\documentclass[preprint,prd,tightenlines,superscriptaddress]{revtex4-1}

\usepackage{bm,amsmath,amsfonts,amssymb,graphicx,xspace,placeins,multirow}

\usepackage{hyperref}
\usepackage{breakurl}
\usepackage{array}
\usepackage{slashed}
\usepackage{siunitx}
\usepackage{xspace}
\usepackage[italic]{hepnames}
\usepackage{booktabs}
\usepackage{hhline}
\usepackage{multirow}
\usepackage{ulem}
\usepackage[]{lineno}
\usepackage{placeins}
\usepackage{lineno}

\newcommand{\bdslnu}{\ensuremath{\Bzb \to D^{*+} \ell^{-} \nub_l}\xspace}

\newcommand{\ifb}{\ensuremath{{\rm fb}^{-1}}\xspace}
\newcommand{\lumi}{\ensuremath{34.6~\ifb}\xspace}
\newcommand{\BR}{{\ensuremath{\cal B}}}
\newcommand{\pis}{{\ensuremath{\pi_s}}\xspace}

\newcommand{\resBF}{\ensuremath{\BR(\bdslnu) = \left(4.51 \pm 0.41_{\mathrm{stat}}\pm0.27_{\mathrm{syst}} \pm0.45_{\mathrm{\pi_s}}\right) \%}\xspace}
\newcommand{\WABF}{\ensuremath{\BR(\bdslnu) = \left(5.05 \pm 0.14 \right) \%}\xspace}

\newcommand{\NBB}{\ensuremath{N_{B \bar B}=  \left(37.7 \pm 0.6 \right) \times 10^6}}

\input ./belle2sym.tex

\begin{document}

\def\belletwo {\it {Belle II}}

\vspace*{-3\baselineskip}
\resizebox{!}{3cm}{\includegraphics{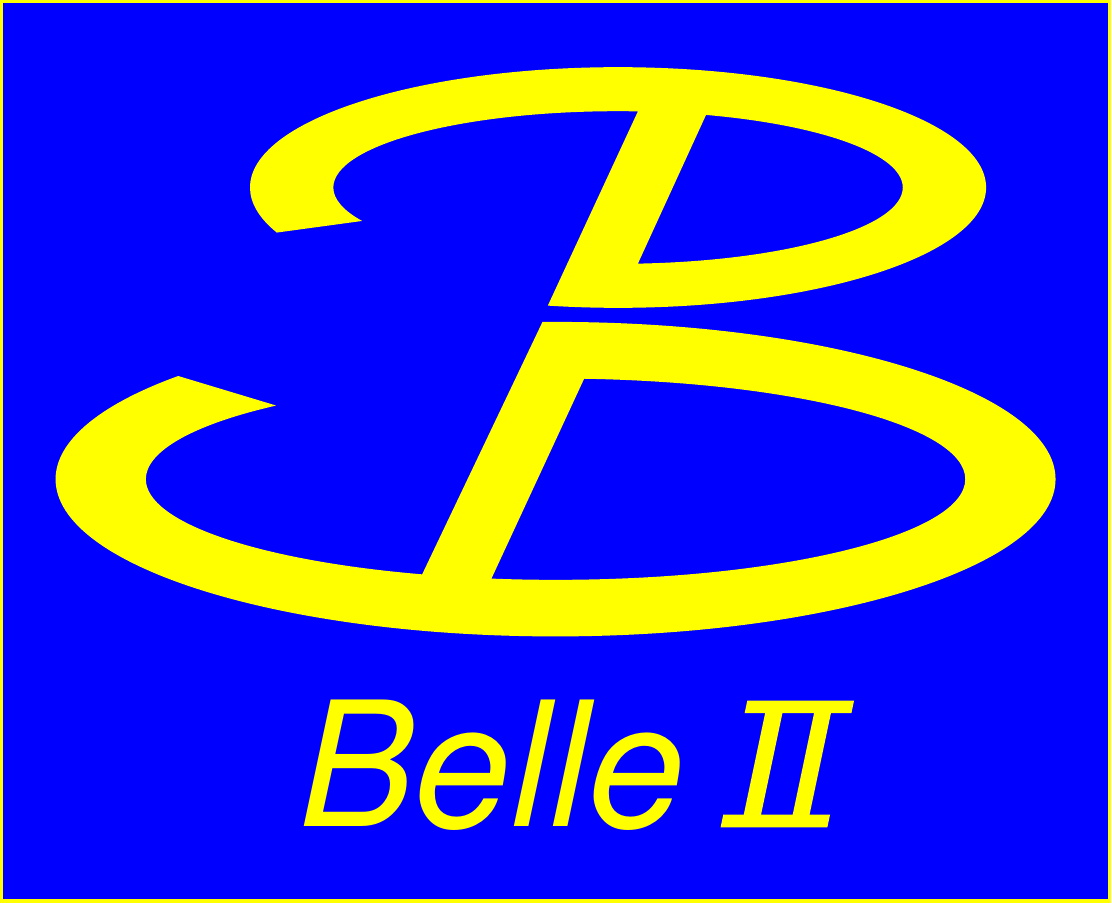}}

\vspace*{-5\baselineskip}
\begin{flushright}
BELLE2-CONF-PH-2020-009\\
\today
\end{flushright}

\title { \quad\\[0.5cm] Measurement of the semileptonic \bdslnu branching fraction with fully reconstructed $B$ meson decays and \mbox{34.6 fb$^{-1}$} of Belle~II data}

\collaboration{The Belle II Collaboration}
\newcommand{\instSinica}{Academia Sinica, Taipei 11529, Taiwan}
\newcommand{\instCPPM}{Aix Marseille Universit\'{e}, CNRS/IN2P3, CPPM, 13288 Marseille, France}
\newcommand{\instBeihang}{Beihang University, Beijing 100191, China}
\newcommand{\instBUAP}{Benemerita Universidad Autonoma de Puebla, Puebla 72570, Mexico}
\newcommand{\instBNL}{Brookhaven National Laboratory, Upton, New York 11973, U.S.A.}
\newcommand{\instBINP}{Budker Institute of Nuclear Physics SB RAS, Novosibirsk 630090, Russian Federation}
\newcommand{\instCMU}{Carnegie Mellon University, Pittsburgh, Pennsylvania 15213, U.S.A.}
\newcommand{\instCinvestavIPN}{Centro de Investigacion y de Estudios Avanzados del Instituto Politecnico Nacional, Mexico City 07360, Mexico}
\newcommand{\instPrague}{Faculty of Mathematics and Physics, Charles University, 121 16 Prague, Czech Republic}
\newcommand{\instChiangMai}{Chiang Mai University, Chiang Mai 50202, Thailand}
\newcommand{\instChiba}{Chiba University, Chiba 263-8522, Japan}
\newcommand{\instChonnam}{Chonnam National University, Gwangju 61186, South Korea}
\newcommand{\instConacyt}{Consejo Nacional de Ciencia y Tecnolog\'{\i}a, Mexico City 03940, Mexico}
\newcommand{\instDESY}{Deutsches Elektronen--Synchrotron, 22607 Hamburg, Germany}
\newcommand{\instDuke}{Duke University, Durham, North Carolina 27708, U.S.A.}
\newcommand{\instITAR}{Institute of Theoretical and Applied Research (ITAR), Duy Tan University, Hanoi 100000, Vietnam}
\newcommand{\instENEA}{ENEA Casaccia, I-00123 Roma, Italy}
\newcommand{\instEri}{Earthquake Research Institute, University of Tokyo, Tokyo 113-0032, Japan}
\newcommand{\instJuelich}{Forschungszentrum J\"{u}lich, 52425 J\"{u}lich, Germany}
\newcommand{\instFuJen}{Department of Physics, Fu Jen Catholic University, Taipei 24205, Taiwan}
\newcommand{\instFudan}{Key Laboratory of Nuclear Physics and Ion-beam Application (MOE) and Institute of Modern Physics, Fudan University, Shanghai 200443, China}
\newcommand{\instGoettingen}{II. Physikalisches Institut, Georg-August-Universit\"{a}t G\"{o}ttingen, 37073 G\"{o}ttingen, Germany}
\newcommand{\instGifu}{Gifu University, Gifu 501-1193, Japan}
\newcommand{\instSOKENDAI}{The Graduate University for Advanced Studies (SOKENDAI), Hayama 240-0193, Japan}
\newcommand{\instGyeongsang}{Gyeongsang National University, Jinju 52828, South Korea}
\newcommand{\instHanyang}{Department of Physics and Institute of Natural Sciences, Hanyang University, Seoul 04763, South Korea}
\newcommand{\instKEK}{High Energy Accelerator Research Organization (KEK), Tsukuba 305-0801, Japan}
\newcommand{\instJPARC}{J-PARC Branch, KEK Theory Center, High Energy Accelerator Research Organization (KEK), Tsukuba 305-0801, Japan}
\newcommand{\instHSE}{Higher School of Economics (HSE), Moscow 101000, Russian Federation}
\newcommand{\instIISER}{Indian Institute of Science Education and Research Mohali, SAS Nagar, 140306, India}
\newcommand{\instIITBhubaneswar}{Indian Institute of Technology Bhubaneswar, Satya Nagar 751007, India}
\newcommand{\instIITGuwahati}{Indian Institute of Technology Guwahati, Assam 781039, India}
\newcommand{\instIITHyderabad}{Indian Institute of Technology Hyderabad, Telangana 502285, India}
\newcommand{\instIITMadras}{Indian Institute of Technology Madras, Chennai 600036, India}
\newcommand{\instIndiana}{Indiana University, Bloomington, Indiana 47408, U.S.A.}
\newcommand{\instIHEPRussia}{Institute for High Energy Physics, Protvino 142281, Russian Federation}
\newcommand{\instHEPHYVienna}{Institute of High Energy Physics, Vienna 1050, Austria}
\newcommand{\instIHEPChina}{Institute of High Energy Physics, Chinese Academy of Sciences, Beijing 100049, China}
\newcommand{\instChennai}{Institute of Mathematical Sciences, Chennai 600113, India}
\newcommand{\instIPP}{Institute of Particle Physics (Canada), Victoria, British Columbia V8W 2Y2, Canada}
\newcommand{\instIOP}{Institute of Physics, Vietnam Academy of Science and Technology (VAST), Hanoi, Vietnam}
\newcommand{\instIFIC}{Instituto de Fisica Corpuscular, Paterna 46980, Spain}
\newcommand{\instFrascati}{INFN Laboratori Nazionali di Frascati, I-00044 Frascati, Italy}
\newcommand{\instNapoliINFN}{INFN Sezione di Napoli, I-80126 Napoli, Italy}
\newcommand{\instPadovaINFN}{INFN Sezione di Padova, I-35131 Padova, Italy}
\newcommand{\instPerugiaINFN}{INFN Sezione di Perugia, I-06123 Perugia, Italy}
\newcommand{\instPisaINFN}{INFN Sezione di Pisa, I-56127 Pisa, Italy}
\newcommand{\instRomaINFN}{INFN Sezione di Roma, I-00185 Roma, Italy}
\newcommand{\instRomaTreINFN}{INFN Sezione di Roma Tre, I-00146 Roma, Italy}
\newcommand{\instTorinoINFN}{INFN Sezione di Torino, I-10125 Torino, Italy}
\newcommand{\instTriesteINFN}{INFN Sezione di Trieste, I-34127 Trieste, Italy}
\newcommand{\instJAEA}{Advanced Science Research Center, Japan Atomic Energy Agency, Naka 319-1195, Japan}
\newcommand{\instMainz}{Johannes Gutenberg-Universit\"{a}t Mainz, Institut f\"{u}r Kernphysik, D-55099 Mainz, Germany}
\newcommand{\instGiessen}{Justus-Liebig-Universit\"{a}t Gie\ss{}en, 35392 Gie\ss{}en, Germany}
\newcommand{\instKarlsruhe}{Institut f\"{u}r Experimentelle Teilchenphysik, Karlsruher Institut f\"{u}r Technologie, 76131 Karlsruhe, Germany}
\newcommand{\instKennesaw}{Kennesaw State University, Kennesaw, Georgia 30144, U.S.A.}
\newcommand{\instKitasato}{Kitasato University, Sagamihara 252-0373, Japan}
\newcommand{\instKISTI}{Korea Institute of Science and Technology Information, Daejeon 34141, South Korea}
\newcommand{\instKorea}{Korea University, Seoul 02841, South Korea}
\newcommand{\instKSU}{Kyoto Sangyo University, Kyoto 603-8555, Japan}
\newcommand{\instKyotoU}{Kyoto University, Kyoto 606-8501, Japan}
\newcommand{\instKyungpook}{Kyungpook National University, Daegu 41566, South Korea}
\newcommand{\instLPI}{P.N. Lebedev Physical Institute of the Russian Academy of Sciences, Moscow 119991, Russian Federation}
\newcommand{\instLNNU}{Liaoning Normal University, Dalian 116029, China}
\newcommand{\instLMU}{Ludwig Maximilians University, 80539 Munich, Germany}
\newcommand{\instLuther}{Luther College, Decorah, Iowa 52101, U.S.A.}
\newcommand{\instMNITJaipur}{Malaviya National Institute of Technology Jaipur, Jaipur 302017, India}
\newcommand{\instMPP}{Max-Planck-Institut f\"{u}r Physik, 80805 M\"{u}nchen, Germany}
\newcommand{\instMPGHLL}{Semiconductor Laboratory of the Max Planck Society, 81739 M\"{u}nchen, Germany}
\newcommand{\instMcGill}{McGill University, Montr\'{e}al, Qu\'{e}bec, H3A 2T8, Canada}
\newcommand{\instMETU}{Middle East Technical University, 06531 Ankara, Turkey}
\newcommand{\instMEPhI}{Moscow Physical Engineering Institute, Moscow 115409, Russian Federation}
\newcommand{\instNagoya}{Graduate School of Science, Nagoya University, Nagoya 464-8602, Japan}
\newcommand{\instNagoyaKMI}{Kobayashi-Maskawa Institute, Nagoya University, Nagoya 464-8602, Japan}
\newcommand{\instNagoyaIAR}{Institute for Advanced Research, Nagoya University, Nagoya 464-8602, Japan}
\newcommand{\instNaraWu}{Nara Women's University, Nara 630-8506, Japan}
\newcommand{\instUNAM}{National Autonomous University of Mexico, Mexico City, Mexico}
\newcommand{\instNTUTaiwan}{Department of Physics, National Taiwan University, Taipei 10617, Taiwan}
\newcommand{\instNUUTaiwan}{National United University, Miao Li 36003, Taiwan}
\newcommand{\instKrakow}{H. Niewodniczanski Institute of Nuclear Physics, Krakow 31-342, Poland}
\newcommand{\instNiigata}{Niigata University, Niigata 950-2181, Japan}
\newcommand{\instNSU}{Novosibirsk State University, Novosibirsk 630090, Russian Federation}
\newcommand{\instOkinawa}{Okinawa Institute of Science and Technology, Okinawa 904-0495, Japan}
\newcommand{\instOsakaCity}{Osaka City University, Osaka 558-8585, Japan}
\newcommand{\instRCNP}{Research Center for Nuclear Physics, Osaka University, Osaka 567-0047, Japan}
\newcommand{\instPNNL}{Pacific Northwest National Laboratory, Richland, Washington 99352, U.S.A.}
\newcommand{\instPanjab}{Panjab University, Chandigarh 160014, India}
\newcommand{\instPeking}{Peking University, Beijing 100871, China}
\newcommand{\instPanjabPAU}{Punjab Agricultural University, Ludhiana 141004, India}
\newcommand{\instRIKENMSL}{Meson Science Laboratory, Cluster for Pioneering Research, RIKEN, Saitama 351-0198, Japan}
\newcommand{\instRIKEN}{Theoretical Research Division, Nishina Center, RIKEN, Saitama 351-0198, Japan}
\newcommand{\instXavier}{St. Francis Xavier University, Antigonish, Nova Scotia, B2G 2W5, Canada}
\newcommand{\instSeoul}{Seoul National University, Seoul 08826, South Korea}
\newcommand{\instShandong}{Shandong University, Jinan 250100, China}
\newcommand{\instSPU}{Showa Pharmaceutical University, Tokyo 194-8543, Japan}
\newcommand{\instSoochow}{Soochow University, Suzhou 215006, China}
\newcommand{\instSoongsil}{Soongsil University, Seoul 06978, South Korea}
\newcommand{\instLjubljanaJSI}{J. Stefan Institute, 1000 Ljubljana, Slovenia}
\newcommand{\instKyiv}{Taras Shevchenko National Univ. of Kiev, Kiev, Ukraine}
\newcommand{\instTata}{Tata Institute of Fundamental Research, Mumbai 400005, India}
\newcommand{\instTUM}{Department of Physics, Technische Universit\"{a}t M\"{u}nchen, 85748 Garching, Germany}
\newcommand{\instECUTUM}{Excellence Cluster Universe, Technische Universit\"{a}t M\"{u}nchen, 85748 Garching, Germany}
\newcommand{\instTelAviv}{Tel Aviv University, School of Physics and Astronomy, Tel Aviv, 69978, Israel}
\newcommand{\instToho}{Toho University, Funabashi 274-8510, Japan}
\newcommand{\instTohoku}{Department of Physics, Tohoku University, Sendai 980-8578, Japan}
\newcommand{\instTitech}{Tokyo Institute of Technology, Tokyo 152-8550, Japan}
\newcommand{\instTokyoMetropolitan}{Tokyo Metropolitan University, Tokyo 192-0397, Japan}
\newcommand{\instUAS}{Universidad Autonoma de Sinaloa, Sinaloa 80000, Mexico}
\newcommand{\instNapoliUNIV}{Dipartimento di Scienze Fisiche, Universit\`{a} di Napoli Federico II, I-80126 Napoli, Italy}
\newcommand{\instNapoliUNIVA}{Dipartimento di Agraria, Universit\`{a} di Napoli Federico II, I-80055 Portici (NA), Italy}
\newcommand{\instPadovaUNIV}{Dipartimento di Fisica e Astronomia, Universit\`{a} di Padova, I-35131 Padova, Italy}
\newcommand{\instPerugiaUNIV}{Dipartimento di Fisica, Universit\`{a} di Perugia, I-06123 Perugia, Italy}
\newcommand{\instPisaUNIV}{Dipartimento di Fisica, Universit\`{a} di Pisa, I-56127 Pisa, Italy}
\newcommand{\instRomaUNIV}{Universit\`{a} di Roma ``La Sapienza,'' I-00185 Roma, Italy}
\newcommand{\instRomaTreUNIV}{Dipartimento di Matematica e Fisica, Universit\`{a} di Roma Tre, I-00146 Roma, Italy}
\newcommand{\instTorinoUNIV}{Dipartimento di Fisica, Universit\`{a} di Torino, I-10125 Torino, Italy}
\newcommand{\instTriesteUNIV}{Dipartimento di Fisica, Universit\`{a} di Trieste, I-34127 Trieste, Italy}
\newcommand{\instMontreal}{Universit\'{e} de Montr\'{e}al, Physique des Particules, Montr\'{e}al, Qu\'{e}bec, H3C 3J7, Canada}
\newcommand{\instIJCLab}{Universit\'{e} Paris-Saclay, CNRS/IN2P3, IJCLab, 91405 Orsay, France}
\newcommand{\instIPHC}{Universit\'{e} de Strasbourg, CNRS, IPHC, UMR 7178, 67037 Strasbourg, France}
\newcommand{\instAdelaide}{Department of Physics, University of Adelaide, Adelaide, South Australia 5005, Australia}
\newcommand{\instBonn}{University of Bonn, 53115 Bonn, Germany}
\newcommand{\instUBC}{University of British Columbia, Vancouver, British Columbia, V6T 1Z1, Canada}
\newcommand{\instCincinnati}{University of Cincinnati, Cincinnati, Ohio 45221, U.S.A.}
\newcommand{\instFlorida}{University of Florida, Gainesville, Florida 32611, U.S.A.}
\newcommand{\instHamburg}{University of Hamburg, 20148 Hamburg, Germany}
\newcommand{\instHawaii}{University of Hawaii, Honolulu, Hawaii 96822, U.S.A.}
\newcommand{\instHeidelberg}{University of Heidelberg, 68131 Mannheim, Germany}
\newcommand{\instLjubljanaUniLJ}{Faculty of Mathematics and Physics, University of Ljubljana, 1000 Ljubljana, Slovenia}
\newcommand{\instLouisville}{University of Louisville, Louisville, Kentucky 40292, U.S.A.}
\newcommand{\instMalaya}{National Centre for Particle Physics, University Malaya, 50603 Kuala Lumpur, Malaysia}
\newcommand{\instLjubljanaUM}{University of Maribor, 2000 Maribor, Slovenia}
\newcommand{\instMelbourne}{School of Physics, University of Melbourne, Victoria 3010, Australia}
\newcommand{\instMississippi}{University of Mississippi, University, Mississippi 38677, U.S.A.}
\newcommand{\instUOM}{University of Miyazaki, Miyazaki 889-2192, Japan}
\newcommand{\instNovaGorica}{University of Nova Gorica, 5000 Nova Gorica, Slovenia}
\newcommand{\instPittsburgh}{University of Pittsburgh, Pittsburgh, Pennsylvania 15260, U.S.A.}
\newcommand{\instUSTC}{University of Science and Technology of China, Hefei 230026, China}
\newcommand{\instSAlabama}{University of South Alabama, Mobile, Alabama 36688, U.S.A.}
\newcommand{\instSCarolina}{University of South Carolina, Columbia, South Carolina 29208, U.S.A.}
\newcommand{\instSydney}{School of Physics, University of Sydney, New South Wales 2006, Australia}
\newcommand{\instTabuk}{Department of Physics, Faculty of Science, University of Tabuk, Tabuk 71451, Saudi Arabia}
\newcommand{\instUTokyo}{Department of Physics, University of Tokyo, Tokyo 113-0033, Japan}
\newcommand{\instIPMU}{Kavli Institute for the Physics and Mathematics of the Universe (WPI), University of Tokyo, Kashiwa 277-8583, Japan}
\newcommand{\instVictoria}{University of Victoria, Victoria, British Columbia, V8W 3P6, Canada}
\newcommand{\instVPI}{Virginia Polytechnic Institute and State University, Blacksburg, Virginia 24061, U.S.A.}
\newcommand{\instWayneState}{Wayne State University, Detroit, Michigan 48202, U.S.A.}
\newcommand{\instYamagata}{Yamagata University, Yamagata 990-8560, Japan}
\newcommand{\instYerevan}{Alikhanyan National Science Laboratory, Yerevan 0036, Armenia}
\newcommand{\instYonsei}{Yonsei University, Seoul 03722, South Korea}
\affiliation{\instCPPM}
\affiliation{\instBeihang}
\affiliation{\instBNL}
\affiliation{\instBINP}
\affiliation{\instCMU}
\affiliation{\instCinvestavIPN}
\affiliation{\instPrague}
\affiliation{\instChiangMai}
\affiliation{\instChiba}
\affiliation{\instChonnam}
\affiliation{\instConacyt}
\affiliation{\instDESY}
\affiliation{\instDuke}
\affiliation{\instITAR}
\affiliation{\instEri}
\affiliation{\instJuelich}
\affiliation{\instFuJen}
\affiliation{\instFudan}
\affiliation{\instGoettingen}
\affiliation{\instGifu}
\affiliation{\instSOKENDAI}
\affiliation{\instGyeongsang}
\affiliation{\instHanyang}
\affiliation{\instKEK}
\affiliation{\instJPARC}
\affiliation{\instHSE}
\affiliation{\instIISER}
\affiliation{\instIITBhubaneswar}
\affiliation{\instIITGuwahati}
\affiliation{\instIITHyderabad}
\affiliation{\instIITMadras}
\affiliation{\instIndiana}
\affiliation{\instIHEPRussia}
\affiliation{\instHEPHYVienna}
\affiliation{\instIHEPChina}
\affiliation{\instIPP}
\affiliation{\instIOP}
\affiliation{\instIFIC}
\affiliation{\instFrascati}
\affiliation{\instNapoliINFN}
\affiliation{\instPadovaINFN}
\affiliation{\instPerugiaINFN}
\affiliation{\instPisaINFN}
\affiliation{\instRomaINFN}
\affiliation{\instRomaTreINFN}
\affiliation{\instTorinoINFN}
\affiliation{\instTriesteINFN}
\affiliation{\instJAEA}
\affiliation{\instMainz}
\affiliation{\instGiessen}
\affiliation{\instKarlsruhe}
\affiliation{\instKitasato}
\affiliation{\instKISTI}
\affiliation{\instKorea}
\affiliation{\instKSU}
\affiliation{\instKyungpook}
\affiliation{\instLPI}
\affiliation{\instLNNU}
\affiliation{\instLMU}
\affiliation{\instLuther}
\affiliation{\instMNITJaipur}
\affiliation{\instMPP}
\affiliation{\instMPGHLL}
\affiliation{\instMcGill}
\affiliation{\instMEPhI}
\affiliation{\instNagoya}
\affiliation{\instNagoyaKMI}
\affiliation{\instNagoyaIAR}
\affiliation{\instNaraWu}
\affiliation{\instNTUTaiwan}
\affiliation{\instNUUTaiwan}
\affiliation{\instKrakow}
\affiliation{\instNiigata}
\affiliation{\instNSU}
\affiliation{\instOkinawa}
\affiliation{\instOsakaCity}
\affiliation{\instRCNP}
\affiliation{\instPNNL}
\affiliation{\instPanjab}
\affiliation{\instPeking}
\affiliation{\instPanjabPAU}
\affiliation{\instRIKENMSL}
\affiliation{\instSeoul}
\affiliation{\instSPU}
\affiliation{\instSoochow}
\affiliation{\instSoongsil}
\affiliation{\instLjubljanaJSI}
\affiliation{\instKyiv}
\affiliation{\instTata}
\affiliation{\instTUM}
\affiliation{\instTelAviv}
\affiliation{\instToho}
\affiliation{\instTohoku}
\affiliation{\instTitech}
\affiliation{\instTokyoMetropolitan}
\affiliation{\instUAS}
\affiliation{\instNapoliUNIV}
\affiliation{\instPadovaUNIV}
\affiliation{\instPerugiaUNIV}
\affiliation{\instPisaUNIV}
\affiliation{\instRomaUNIV}
\affiliation{\instRomaTreUNIV}
\affiliation{\instTorinoUNIV}
\affiliation{\instTriesteUNIV}
\affiliation{\instMontreal}
\affiliation{\instIJCLab}
\affiliation{\instIPHC}
\affiliation{\instAdelaide}
\affiliation{\instBonn}
\affiliation{\instUBC}
\affiliation{\instCincinnati}
\affiliation{\instFlorida}
\affiliation{\instHawaii}
\affiliation{\instHeidelberg}
\affiliation{\instLjubljanaUniLJ}
\affiliation{\instLouisville}
\affiliation{\instMalaya}
\affiliation{\instLjubljanaUM}
\affiliation{\instMelbourne}
\affiliation{\instMississippi}
\affiliation{\instUOM}
\affiliation{\instPittsburgh}
\affiliation{\instUSTC}
\affiliation{\instSAlabama}
\affiliation{\instSCarolina}
\affiliation{\instSydney}
\affiliation{\instUTokyo}
\affiliation{\instIPMU}
\affiliation{\instVictoria}
\affiliation{\instVPI}
\affiliation{\instWayneState}
\affiliation{\instYamagata}
\affiliation{\instYerevan}
\affiliation{\instYonsei}
  \author{F.~Abudin{\'e}n}\affiliation{\instTriesteINFN} 
  \author{I.~Adachi}\affiliation{\instKEK}\affiliation{\instSOKENDAI} 
  \author{R.~Adak}\affiliation{\instFudan} 
  \author{K.~Adamczyk}\affiliation{\instKrakow} 
  \author{P.~Ahlburg}\affiliation{\instBonn} 
  \author{J.~K.~Ahn}\affiliation{\instKorea} 
  \author{H.~Aihara}\affiliation{\instUTokyo} 
  \author{N.~Akopov}\affiliation{\instYerevan} 
  \author{A.~Aloisio}\affiliation{\instNapoliUNIV}\affiliation{\instNapoliINFN} 
  \author{F.~Ameli}\affiliation{\instRomaINFN} 
% ...
\collaboration{Belle II Collaboration}

\begin{abstract}

We present a first measurement of the \bdslnu branching fraction using fully reconstructed $B$ meson decays employing the Full Event Interpretation algorithm. Collision events corresponding to an integrated luminosity of \lumi are analyzed, which were recorded by the Belle~II detector operated at the SuperKEKB accelerator complex. We measure \mbox{\resBF}, with the first and second error denoting the statistical and systematic uncertainty, respectively, and the third dominant uncertainty is from the slow pion reconstruction efficiency. 

\end{abstract}


\maketitle

{\renewcommand{\thefootnote}{\fnsymbol{footnote}}}
\setcounter{footnote}{0}

\section{Introduction}

Precision measurements of semileptonic \bdslnu decays are crucial for future measurements of the absolute value of the Cabibbo-Kobayashi-Maskawa matrix element $V_{cb}$. A very clean measurement approach for this final state is to fully reconstruct one of the two $B$ mesons produced in the $e^+ e^- \to \Upsilon(4S) \to B \overline B$ process, using hadronic modes. The flavor and momentum of the signal $B$ meson can thus be determined. In this conference note, we present a first study of \bdslnu decays with fully reconstructed $B$ meson events using the Full Event Interpretation (FEI) algorithm of Ref.~\cite{Keck:2018lcd}. The FEI reconstructs over 100 $B$ meson decay channels and over 10,000 decay cascades. 

\section{The Belle~II detector and data sample}

The Belle~II detector~\cite{Abe:2010sj, ref:b2tip} operates at the SuperKEKB asymmetric-energy  electron-positron collider~\cite{superkekb}, located at the KEK laboratory in Tsukuba, Japan. The detector consists of several nested detector subsystems arranged around the beam pipe in a cylindrical geometry. The innermost subsystem is the vertex detector, which includes two layers of silicon pixel detectors and four outer layers of silicon strip detectors. Currently, the second pixel layer is installed in only a small part of the solid angle, while the remaining vertex detector layers are fully installed. Most of the tracking volume consists of a helium and ethane-based small-cell drift chamber (CDC). Outside the drift chamber, a Cherenkov-light imaging and time-of-propagation detector provides charged-particle identification in the barrel region. In the forward endcap, this function is provided by a proximity-focusing, ring-imaging Cherenkov detector with an aerogel radiator. Further out is an electromagnetic calorimeter, consisting of a barrel and two endcap sections made of CsI(Tl) crystals. A uniform 1.5~T magnetic field is provided by a superconducting solenoid situated outside the calorimeter. Multiple layers of scintillators and resistive plate chambers, located between the magnetic flux-return iron plates, constitute the $K_L$ and muon identification system.

The data used in this analysis were collected in 2019 and 2020 at a center-of-mass (CM) energy of 10.58~GeV, corresponding to the mass of the $\Upsilon$(4S) resonance. The energies of the electron and positron beams are $7\gev$ and $4\gev$, respectively, resulting in a boost of $\beta\gamma = 0.28$ of the CM frame relative to the laboratory frame. The number of $B$ meson pairs in the analyzed collision events has been counted using event-shape variables and has been determined to be \NBB.

Simulated Monte Carlo (MC) samples are used to develop the signal selection, determine reconstruction efficiencies and understand potential background distributions. These samples are generated using EvtGen and consist of generic \BB events where  $e^+ e^-\rightarrow \Upsilon(4S) \rightarrow B\bar B$ and $e^+e^-\to q\bar q~ (q = u, d, s, c)$. The latter is simulated with KKMC~\cite{Ward:2002qq} and PYTHIA~\cite{Sjostrand:2007gs}. The corresponding luminosity of the generic and continuum samples is \mbox{100 \invfb}. A signal MC sample, where $B$ exclusively decays to $ X_{c} \ell \nu_{\ell}$, is also generated and used in this analysis. All recorded collisions and simulated events are analyzed in the basf2~\cite{basf2} framework. Data-driven corrections for the lepton identification are applied to the MC events, derived from $J/\psi$ and other control samples.

\section{Full Event Interpretation and Event selection}

We first reconstruct collision events using the FEI algorithm.  The algorithm reconstructs one of the \PB mesons produced in the collision event using hadronic decay channels. We label such \PB mesons in the following as $B_{\mathrm{tag}}$. Instead of attempting to reconstruct as many \PB meson decay cascades as possible, the algorithm employs a hierarchical reconstruction ansatz in six stages: in the first stage, tracks, displaced vertices and neutral clusters are identified and required to pass some basic quality criteria. In the second stage, boosted decision trees (BDTs) are trained to identify charged tracks and neutral energy depositions as detector stable particles ($e^+, \mu^+, K^+, K_L, p, \pi^+, \gamma$). In the third and fourth stage, these candidate particles are combined into composite parents ($J/\Psi, D^0, D^{+}, D_s, \Lambda_c, \Lambda, \Sigma^+$), and for each target final state, a BDT is trained to identify probable candidates. At the fifth stage, candidates for excited mesons ($D^{*\, 0}, D^{*\, +}, D_s^*$) are formed and separate BDTs are trained to identify viable combinations. The input variables of each stage aggregate the output classifiers from all previous reconstruction stages. The final stage combines the information from all previous stages to form $B_{\mathrm{tag}}$ candidates. The viability of such combinations is again assessed by a BDT that is trained to distinguish correctly reconstructed candidates from wrong combinations and whose output classifier score we denote as signal probability. We apply a calibration factor for the hadronic tagging efficiency on MC derived using inclusive $B \to X \ell \bar \nu_\ell$. A full description of this procedure can be found in Ref.~\cite{Sutcliffe:1992}. 

Only events with at least three charged tracks and three neutral clusters are passed into the FEI algorithm. The distance of closest approach between each track and the interaction point must be less than 2 \cm and 0.5\cm along and longitudinal to the beam axis, respectively, with a minimum transverse momentum of 100 \mevc. Clusters must have an energy of at least 100 \mev and the associated polar angle is required to lie within the angular acceptance of the CDC, $\ang{17 } < \theta < \ang{150}$, with $\theta$ denoting the polar angle in the laboratory frame. In addition, to exclude low multiplicity events such as $\epem \to\epem$ from the FEI reconstruction, the following selection is applied: $2<E_{ECL}<7 \gev$ and $E_{vis}>4 \gev$. The former is the total energy deposited  in the electromagnetic calorimeter and the latter is determined using the energy of all the tracks and clusters in the event. Collision events where $\epem\rightarrow q\bar q$ ($q = u,\,d,\,s,\,c$) are also suppressed by demanding $R_2 <0.3$, where $R_2$ is the ratio of the second and zeroth Fox-Wolfram moments~\cite{Fox:1978vu}, calculated using all the tracks and photon candidates in the event.

The purity of the $B_{\mathrm{tag}}$ candidate is improved by selecting only candidates with an output signal probability greater than $0.001$. The beam-constrained mass $M_{\rm{bc}}$ of a $B_{\mathrm{tag}}$ candidate is defined as
\begin{equation}
    M_{\rm{bc}}=\sqrt{E^2_{CM}- |\vec{p}_{B_{\mathrm{tag}}}|^2}
\end{equation}
where $E_{CM}$ is half the total collision energy, employed here to avoid resolution uncertainties related to the measurement of the $B$ energy, and $\vec{p}_{B_{\mathrm{tag}}}$ is the momentum of the $B_{\mathrm{tag}}$ candidate in the CM frame. $B_{\mathrm{tag}}$ candidates are required to have $m_{bc}$ greater than $5.27 \gevcc$ and $\Delta E \in [-0.15,0.1]$\gev, where $\Delta E =E_{B_{\mathrm{tag}}} - E_{CM}s$, with $E_{B_{\mathrm{tag}}}$ denoting the CM frame energy of the $B_{\mathrm{tag}}$ candidate.

All tracks and neutral clusters used in the $B_{\mathrm{tag}}$  reconstruction are masked in the event. All remaining tracks and clusters are then used to define the signal side. The decay cascade $B^0 \rightarrow \Dstarp\ell\nu_{\ell}$ with $\Dstarp \rightarrow \Dz \pip$ and $\Dz \rightarrow K^- \pip$ (and charge conjugate) is reconstructed to form a $B_{\mathrm{sig}}$ candidate. All tracks must fulfill the same quality criteria as described above and, except for the slow pion $\pi_s$ daughter produced in the \Dstarp decay, must have at least one hit in the CDC. Oppositely charged tracks are combined to form \Dz candidates. Each \Dz meson candidate is required to have an invariant mass conforming to $m_{K\pi} \in [1.858, 1.878] \gevcc$ and a CM momentum of less than 3 \gevc. The  \Dz meson candidates are then combined with a third track to form the \Dstarp candidate. The mass difference, defined as $\Delta{m} = m_{D^{*}}-m_{D}$, must lie within $[0.143, 0.148] \gevcc$.
In addition, charged leptons must pass lepton particle identification (PID) criteria in the form of a likelihood, which is determined using information from the different detector subsystems and ranges from zero to unity. Each lepton candidate must have a PID likelihood ratio greater than 0.9 to be selected as either an electron or muon candidate. In addition, we require that lepton candidates have a CM momentum greater than 1 \gevc. 
The lepton candidate is then combined with an oppositely charged $D^*$ candidate and constrained with a vertex fit, requiring both daughters to originate from a common point. $\FourS$ candidates are formed by combining the resulting $D^*\ell$ candidate with a $B_{\mathrm{tag}}$. Events with additional tracks, after the $\FourS$ reconstruction, are excluded. At this point, there are on average 1.4 \FourS candidates per event. We select the candidate with the highest FEI signal probability of the daughter $B_{\mathrm{tag}}$. If an event still has more than one candidate per event (which occurs for about 1.8\% of all remaining events), we select the candidate with its $D^*$ candidate mass closest to the world average $D^*$ mass. To reduce possible backgrounds from fully hadronic decays, we also impose that the missing energy, $E_{\mathrm{miss}} =  2 \times E_{CM} - E_{B_{\mathrm{tag}}} - E_{D^*} - E_\ell$, exceeds 300 MeV. Here $E_{D^*}$ and $E_\ell$ denote the energy of the reconstructed $D^*$ and lepton candidates, and is calculated using the reconstructed momenta.

\section{Signal Extraction and branching fraction}

\begin{figure}
\begin{center}
\includegraphics[width=0.75\columnwidth]{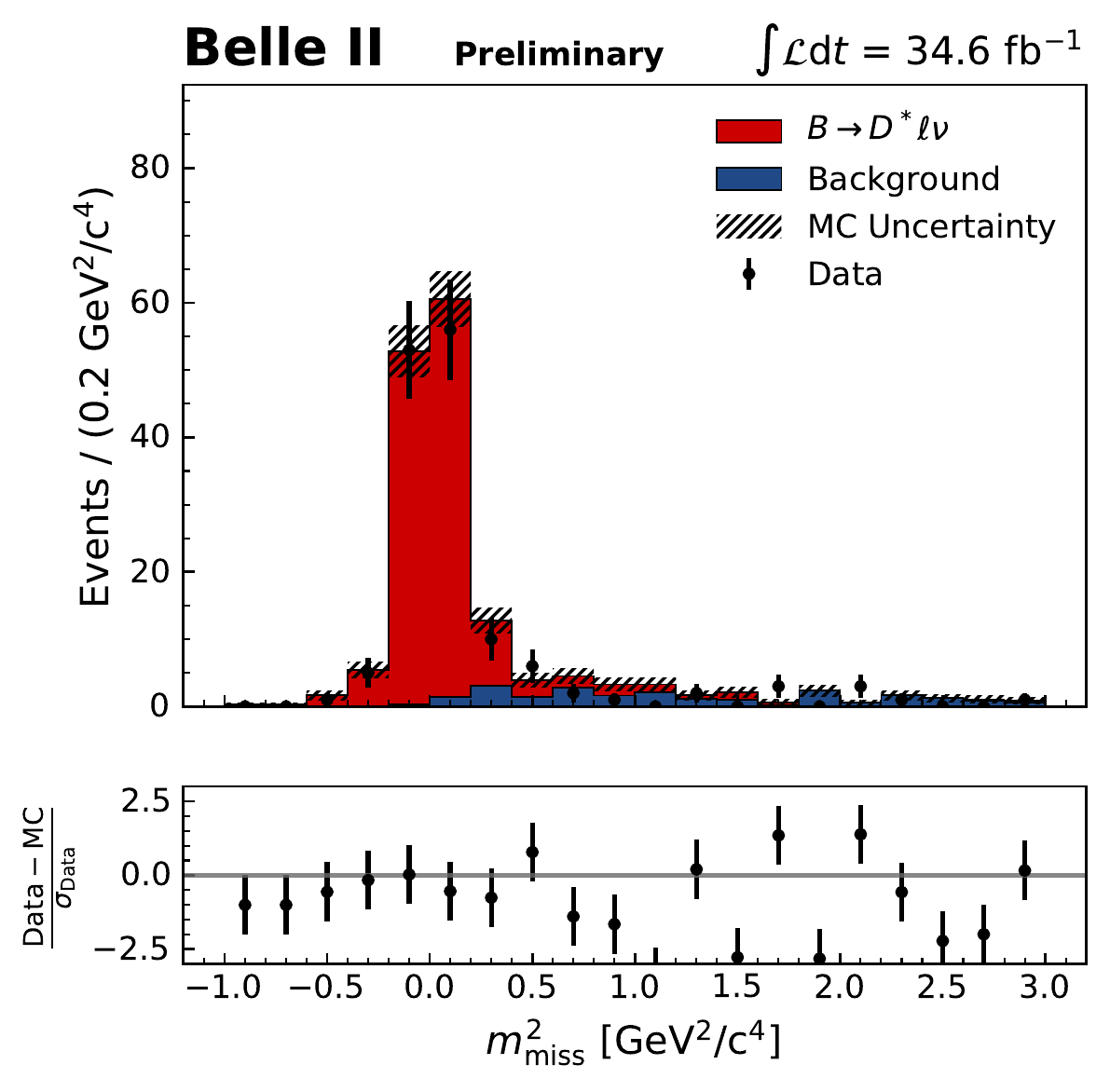} 
  \caption{The reconstructed pre-fit $m_{\mathrm{miss}}^2$ distribution is shown and compared to the MC expectation. The resolution of the peak is dominated by the resolution of the $B_{\mathrm{tag}}$ reconstruction. Correctly reconstructed $B_{\mathrm{sig}}$ candidates are expected to peak at $m_{\mathrm{miss}}^2 \approx m_\nu^2 \sim 0$.
 }
  \label{fig:pre_fit_results}
\end{center}
\end{figure}

The signal is extracted using a binned maximum likelihood fit of $m_{\mathrm{miss}}^2$, defined as
\begin{align}
  m_{\mathrm{miss}}^2 = \left( p_{e^+\, e^-} - p_{B_{\mathrm{tag}}} - p_{D^*} - p_\ell \right)^2 \, ,
\end{align}
and evaluated in the CM frame with $ p_{e^+\, e^-} $ and  $p_{B_{\mathrm{tag}}} = ( E_{CM},  \vec{p}_{B_{\mathrm{tag}}})$ denoting the four-momenta of the colliding electron-positron pair and the reconstructed $B_{\mathrm{tag}}$ candidate. Further, $p_\ell$ and $ p_{D^*} $ denote the four-momenta of the reconstructed lepton and $D^{*\,+}$ candidate. Correctly reconstructed \bdslnu events should peak close to $m_{\mathrm{miss}}^2 \approx m_\nu^2 \sim 0$, whereas contributions from most background processes will have on average larger values. The $m_{\mathrm{miss}}^2 $ distribution of the reconstructed candidate events is shown in Figure~\ref{fig:pre_fit_results}. For the fit we merge the small background contributions from continuum processes and other $B$ meson decays. 

\begin{figure}
\begin{center}
\includegraphics[width=0.75\columnwidth]{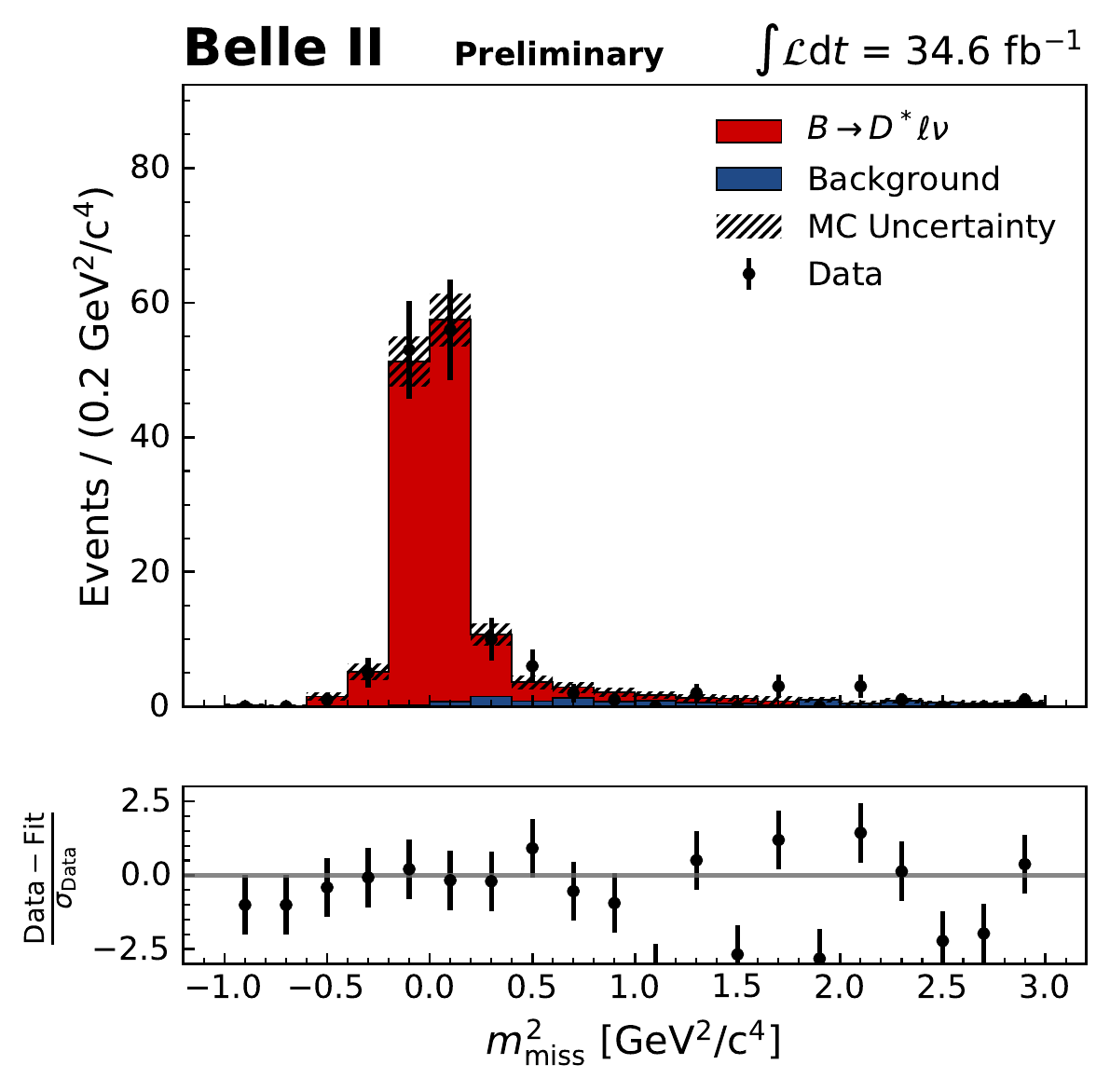} 
  \caption{The post-fit $m_{\mathrm{miss}}^2$ distribution is shown.}
  \label{fig:post_fit_results}
\end{center}
\end{figure}

The fit finds $N_s = 133 \pm 12$ signal and $11 \pm 5$ background events and the fit result is shown in Figure~\ref{fig:post_fit_results}. The fitted yields can be converted into a branching fraction using 
\begin{align}
 \BR(\bdslnu)=  \frac{N_s \times \epsilon^{-1}_{\mathrm{tag+sel}}}{ 4 \times N_{B\overline B} \times \left( 1 + f_{+0}\right)^{-1} } \, .
\label{eq:BR-eq}
\end{align} 
Here $\epsilon_{\mathrm{tag+sel}} = \left( 0.40 \pm 0.05  \right) \times 10^{-4}$ denotes the selection and tagging efficiencies including sub decay branching fractions. The quoted error includes uncertainties from the tagging calibration, the limited size of the MC sample, the lepton identification, the slow pion reconstruction, tracking efficiency, and from the assumed charm branching fractions. Using the preliminary $B$ counting result of \NBB\ and $f_{+0} = 1.058 \pm 0.024$ from Ref.~\cite{pdg:2020} we obtain
\begin{align}
 \resBF \, .
\end{align}
The largest uncertainty stems from the slow pion efficiency and a detailed breakdown is given in Table~\ref{tab:syst}. The measured value is lower, but in good agreement with the world average of Ref.~\cite{pdg:2020} of \WABF.

\begin{table}[htbp]
\begin{center}
\begin{tabular}{l|c}\hline\hline
Source & Relative uncertainty (\%) \\
\hline
Tracking of $\pis$ & 10\%  \\
MC modeling & 5\%\\
FEI Calibration & 3\% \\
Tracking of $K$, $\pi$, $\ell$ & 3\%  \\
$N_{B^0}$ & 2\% \\
$f_{+0}$ & 1\% \\
Charm branching fractions & 1\% \\
Lepton ID & 1\%\\
\hline
Total & 12\% \\
 \hline\hline
\end{tabular}
\end{center}
\caption{Summary of the relative systematic uncertainties for the branching fraction measurement.
}\label{tab:syst}
\end{table}

 \section{$E_{\mathrm{ECL}}$ of the selected \bdslnu events}

\begin{figure}
\begin{center}
\includegraphics[width=0.49\columnwidth]{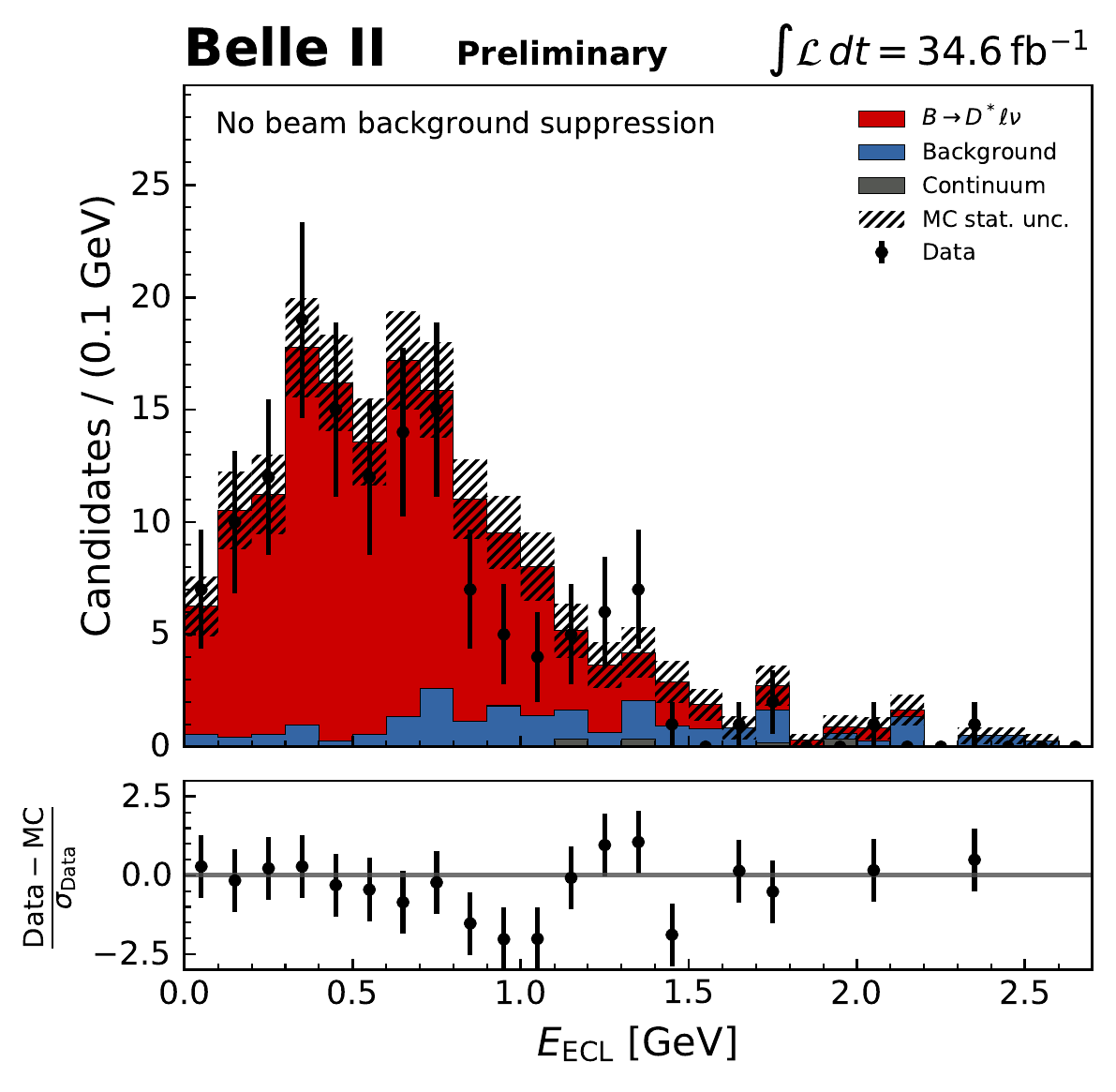} 
\includegraphics[width=0.49\columnwidth]{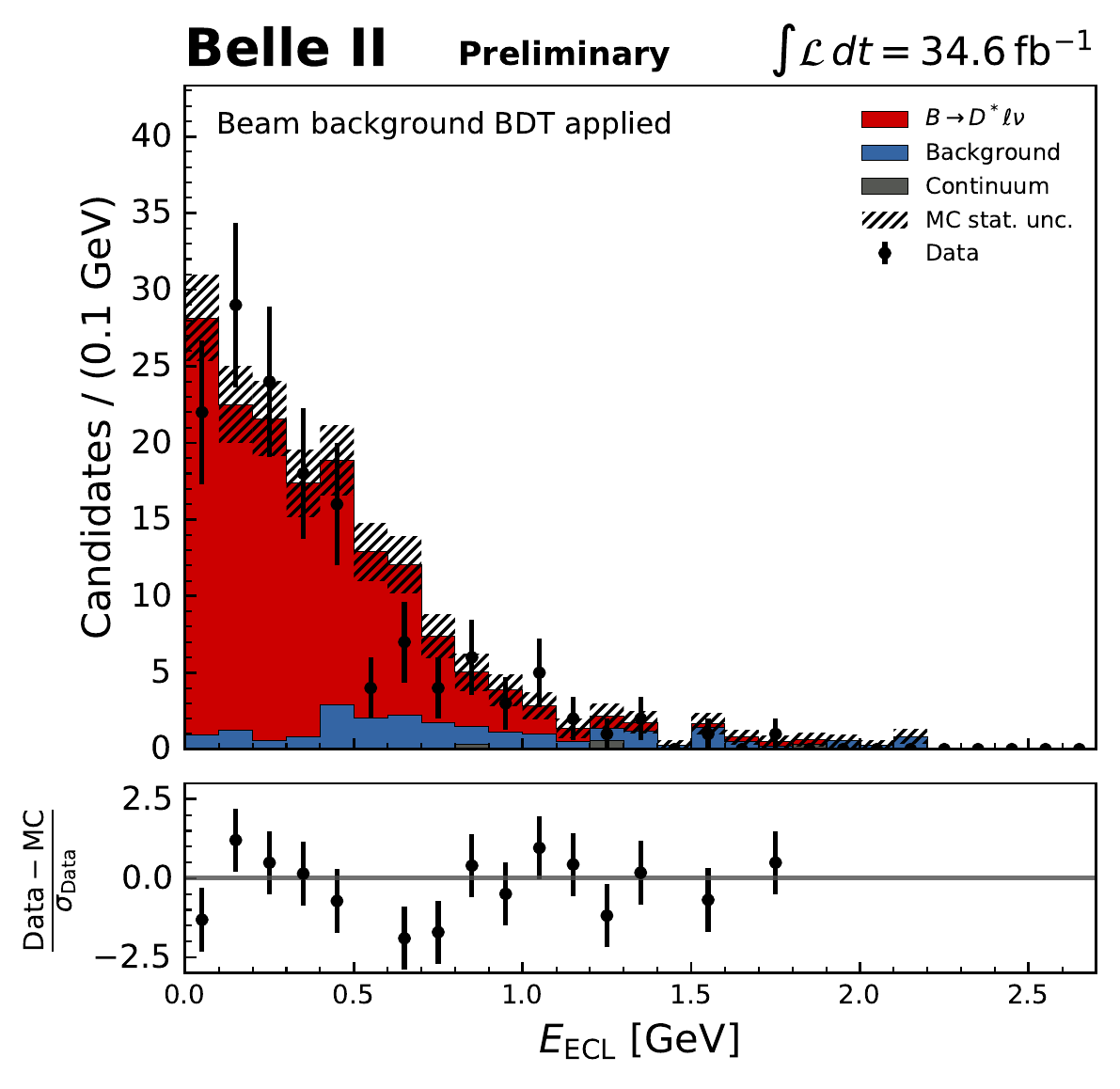} 
  \caption{Two versions of $E_{\mathrm{ECL}}$ are shown: (left) is the version applying detector region dependent energy selection criteria, (right) shows the impact of using a BDT to identify neutral energy depositions from beam background processes. It is based on shower shape variables and the detector region of the reconstructed neutral cluster.}
  \label{fig:ecl_figures}
\end{center}
\end{figure}

The full reconstruction of $B_{\mathrm{tag}}$ and $B_{\mathrm{sig}}$ allows one to analyze unassigned energy depositions in the calorimeter. Their energy can be summed, after some minimal energy cuts and are denoted as $E_{\mathrm{ECL}}$. For correctly reconstructed $B_{\mathrm{sig}}$ candidates, no unassigned neutral energy clusters are expected in the rest of the event (ROE) after the \FourS reconstruction and thus ideally $E_{\mathrm{ECL}} \sim 0$.
Figure~\ref{fig:ecl_figures} left shows $E_{\mathrm{ECL}}$, where only  neutral cluster with energy greater than 60, 30, and 90 \mev in the forward, barrel and end-cap regions of the calorimeter, respectively,  are considered. The resulting distribution for signal events has a tail towards larger values due to unassigned $K_L$ and beam background photons. 

To suppress contributions from beam background photons, a boosted decision tree (BDT) (using the implementation of Ref.~\cite{Keck:BDT}) is trained using 6 variables related to the shape of the electromagnetic shower in the ECL These include the ratio of the energy of the central crystal in a cluster to the summed energy of the 9x9 surrounding crystals, the lateral energy distribution of a given cluster, the second moment of the cluster's energy distribution, the polar and azimuthal angle of each cluster in the ECL, and the output of a multivariate trained on eleven Zernike moments of the cluster shower~\cite{zernike}. The classifier is trained using recorded events in a control sample, where $\epem\rightarrow\mu^+ \mu^-$ with the requirement that the two muons are back to back. The clusters in the control sample result mainly from beam background photons and thus are ideal for training the classifier. 

The classifier is then applied to the clusters of the $E_{\mathrm{ECL}}$ distribution from \bdslnu signal events to evaluate their likeness to beam background photons. A loose cut is applied to exclude clusters that are most likely from beam backgrounds and the resulting $E_{\mathrm{ECL}}$ is shown in Figure~\ref{fig:ecl_figures}. Both  $E_{\mathrm{ECL}}$ distributions, before and after applying BDT selection, show good agreement within the available event counts.  $E_{\mathrm{ECL}}$ is a key experimental observable to measure semileptonic or leptonic $B$ meson decays involving $\tau$ leptons.

\section{Results and Conclusion}

We present a first measurement of the \bdslnu branching fraction using the Full Event Interpretation algorithm and \lumi of Belle~II data. We determine
\begin{align}
 \resBF \, ,
\end{align}
which is lower than, but in agreement with, the current world average. The largest systematic uncertainty stems from the slow pion efficiency, which will be improved in the future with more precise auxiliary measurements. For future studies of $B \to \tau \bar \nu_\tau$ and $B \to D^{(*)} \, \tau \bar \nu_\tau$ from Belle~II, we have also looked at $E_{\mathrm{ECL}}$, defined as the sum of unassigned neutral energy in the calorimeter. The results of these studies are an important stepping stone for future measurements involving these challenging signatures.

\section{ACKNOWLEDGEMENTS}
We thank the SuperKEKB group for the excellent operation of the
accelerator; the KEK cryogenics group for the efficient
operation of the solenoid; and the KEK computer group for
on-site computing support.
This work was supported by the following funding sources:
Science Committee of the Republic of Armenia Grant No. 18T-1C180;
Australian Research Council and research grant Nos.
DP180102629, 
DP170102389, 
DP170102204, 
DP150103061, 
FT130100303, 
and
FT130100018; 
Austrian Federal Ministry of Education, Science and Research, and
Austrian Science Fund No. P 31361-N36; 
Natural Sciences and Engineering Research Council of Canada, Compute Canada and CANARIE;
Chinese Academy of Sciences and research grant No. QYZDJ-SSW-SLH011,
National Natural Science Foundation of China and research grant Nos.
11521505,
11575017,
11675166,
11761141009,
11705209,
and
11975076,
LiaoNing Revitalization Talents Program under contract No. XLYC1807135,
Shanghai Municipal Science and Technology Committee under contract No. 19ZR1403000,
Shanghai Pujiang Program under Grant No. 18PJ1401000,
and the CAS Center for Excellence in Particle Physics (CCEPP);
the Ministry of Education, Youth and Sports of the Czech Republic under Contract No.~LTT17020 and 
Charles University grants SVV 260448 and GAUK 404316;
European Research Council, 7th Framework PIEF-GA-2013-622527, 
Horizon 2020 Marie Sklodowska-Curie grant agreement No. 700525 `NIOBE,' 
and
Horizon 2020 Marie Sklodowska-Curie RISE project JENNIFER2 grant agreement No. 822070 (European grants);
L'Institut National de Physique Nucl\'{e}aire et de Physique des Particules (IN2P3) du CNRS (France);
BMBF, DFG, HGF, MPG, AvH Foundation, and Deutsche Forschungsgemeinschaft (DFG) under Germany's Excellence Strategy -- EXC2121 ``Quantum Universe''' -- 390833306 (Germany);
Department of Atomic Energy and Department of Science and Technology (India);
Israel Science Foundation grant No. 2476/17
and
United States-Israel Binational Science Foundation grant No. 2016113;
Istituto Nazionale di Fisica Nucleare and the research grants BELLE2;
Japan Society for the Promotion of Science,  Grant-in-Aid for Scientific Research grant Nos.
16H03968, 
16H03993, 
16H06492,
16K05323, 
17H01133, 
17H05405, 
18K03621, 
18H03710, 
18H05226,
19H00682, 
26220706,
and
26400255,
the National Institute of Informatics, and Science Information NETwork 5 (SINET5), 
and
the Ministry of Education, Culture, Sports, Science, and Technology (MEXT) of Japan;  
National Research Foundation (NRF) of Korea Grant Nos.
2016R1\-D1A1B\-01010135,
2016R1\-D1A1B\-02012900,
2018R1\-A2B\-3003643,
2018R1\-A6A1A\-06024970,
2018R1\-D1A1B\-07047294,
2019K1\-A3A7A\-09033840,
and
2019R1\-I1A3A\-01058933,
Radiation Science Research Institute,
Foreign Large-size Research Facility Application Supporting project,
the Global Science Experimental Data Hub Center of the Korea Institute of Science and Technology Information
and
KREONET/GLORIAD;
Universiti Malaya RU grant, Akademi Sains Malaysia and Ministry of Education Malaysia;
Frontiers of Science Program contracts
FOINS-296,
CB-221329,
CB-236394,
CB-254409,
and
CB-180023, and SEP-CINVESTAV research grant 237 (Mexico);
the Polish Ministry of Science and Higher Education and the National Science Center;
the Ministry of Science and Higher Education of the Russian Federation,
Agreement 14.W03.31.0026;
University of Tabuk research grants
S-1440-0321, S-0256-1438, and S-0280-1439 (Saudi Arabia);
Slovenian Research Agency and research grant Nos.
J1-9124
and
P1-0135; 
Agencia Estatal de Investigacion, Spain grant Nos.
FPA2014-55613-P
and
FPA2017-84445-P,
and
CIDEGENT/2018/020 of Generalitat Valenciana;
Ministry of Science and Technology and research grant Nos.
MOST106-2112-M-002-005-MY3
and
MOST107-2119-M-002-035-MY3, 
and the Ministry of Education (Taiwan);
Thailand Center of Excellence in Physics;
TUBITAK ULAKBIM (Turkey);
Ministry of Education and Science of Ukraine;
the US National Science Foundation and research grant Nos.
PHY-1807007 
and
PHY-1913789, 
and the US Department of Energy and research grant Nos.
DE-AC06-76RLO1830, 
DE-SC0007983, 
DE-SC0009824, 
DE-SC0009973, 
DE-SC0010073, 
DE-SC0010118, 
DE-SC0010504, 
DE-SC0011784, 
DE-SC0012704; 
and
the National Foundation for Science and Technology Development (NAFOSTED) 
of Vietnam under contract No 103.99-2018.45.

\bibliography{conf.bib}

\end{document}